# Is the San Andreas Fracture a bayonet-shaped fracture as inferred from the acoustic body waves in the SAFOD Pilot hole ?

André Rousseau
*CNRS-UMS 2567 (OASU)*
*Université Bordeaux 1 - Groupe d'Etude des Ondes en Géosciences*
*351, cours de la Libération,F-33405 Talence cédex*
a.rousseau@geog.u-bordeaux1.fr

**Summary**
　　The method using the propagation of acoustic body waves within the stress modified areas around a vertical borehole has been applied to the granitic formation penetrated by the SAFOD Pilot hole near the San Andreas Fault trace. This method allows us investigating the horizontal in situ stresses.
　　Only P waves supplied useful and surprising information. A depth of 1270 m separates an upper region of uniform thickness of stress modified areas, possibly corresponding to a shear domain, and a lower region where there are simultaneously two values of the thicknesses of the stress modified areas (particularly between 1500 and 1600 m of depth) possibly corresponding to a compressive and a shear domain.
　　In order to integrate the contradictory effects of the simultaneity of shear and compressive domains at some depths, as well as the presence of three shear zones at particular depths, we propose that the San Andreas Fault could be bayonet-shaped instead of planar. Other recent available information in the literature about this fault, such as the presence of a fault zone of low shear wave velocity, stress rotation measured with depth, and the large angles of the frictional coefficients, can be logically explained by this kind of fault geometry.

**Introduction**
　　The method used to analyse the propagation of acoustic body waves within the stress modified areas around a vertical borehole in order to investigate the horizontal *in situ* stresses (see in Rousseau [2005] the principles of this method) has been applied to data from the SAFOD Pilot Hole drilled near Parkfield in California, close to the surface trace of the San Andreas Fault. This model of contemporary stress is relevant because it was applied to the part of the well drilled within crystalline rocks, the Salinian granite.
　　The log data obtained in this well have been interpreted by Boness and Zoback (2004) who show the correlation between stress-induced seismic velocity anisotropy and physical properties from the flexural waves generated by a dipole tool. In this paper we investigate the behaviour of the acoustic body waves generated by a monopole source. After characterising the double P and S waves and calculating their velocities, we will evaluate the radial thicknesses of the stress deformed areas from the multiples propagating inside them, and propose an adequate model of fracture.

**Consideration about the body waves from the SAFOD Pilot Hole**
　　Figure 1 displays in 2D an example - between fractures - of the move-out of the body waves of frequency 13.3 kHz generated by a shot over eight receivers spaced by a half foot. Well individualized on each trace are two P waves, P1 and P2, and up to three S waves, S1,

S2 and S3. P2 is considered as a multiple of P1 inside the stress modified area, and S2 and S3 a multiple of S1 inside the same area.

The mean velocities of P1, P2, S1 and S2 have been calculated from their move-out using the "sequential" method without interpolation (see Rousseau [2005]). They are represented on Figure 2. The results have been plotted on this diagram only if the eight receivers had intervened in the calculation and the standard deviation been inferior to 50 %.
- The mean velocities of P waves vary between 5100 and 6200 m/s, except between 1100 and 1450 meters of depth where the range is rather 5100-5800 m/s. Below 1900 meters of depth, their velocity tends to increase.
- The range of the mean velocities of S waves is narrower than that of P waves, but we can observe undulations in respect to depth. The velocities are comprised between 2500 m/s and 3500 m/s, and are faster than 3000 m/s below 1900 meters of depth.
- The mean velocities of the multiples do not differ a great deal from those of the first arrivals.
- The depths without values correspond to the shear zones : the software did not keep the values which were unreliable as a consequence of the method used.

**Thicknesses of the stress modified areas around the hole**

The thicknesses of the stress modified areas calculated from the mean velocities of P2 and S2 and the times spent (i) between P1 and P2 concerning P waves, (ii) S1 and S2 concerning S waves, are plotted on Figure 3. If the results yielded by P waves appear organized, those by S waves are scattered, and, as a consequence, probably not representative. This may be due to the presence of breakouts, which disturb the shear waves more than the pressure waves because of their vertical direction.

Figure 4 shows, for the results calculated from P waves (Figure 4a) and those calculated from S waves (Figure 4b) : 1) the distribution of the thicknesses over 50 meters of depth in respect to depth by grades of 0.1 meter, the width between two vertical lines representing 100 % of the corresponding size, 2) the percent of the "normal velocities" (the first arrival is faster than the second one), and 3) the percent of the "inverted velocities" (the first arrival is slower than the second one). The values of the thicknesses yielded by P waves supply fruitful information :
- there is a sharp difference above and below 1270 meters of depth,
- above this depth, the thicknesses of the modified areas are mainly of 0.4 meter,
- below this depth, there are simultaneously two values of those thicknesses : 0.4 meter and between 0.5 and 0.6 meter, particularly between 1500-1600 meters of depth,
- a noticeable proportion of "inverted" velocities characterizes the modified areas.

**Stress models from the modified areas thicknesses**

The model of horizontal stresses has been calculated as indicated in Rousseau (2005). The goal is to fit probable horizontal stresses to the thickness value of the calculated stress modified area at a given depth in relation to the wavelength (see Figure 5). The results obtained only by P waves have been taken into account : as their velocity oscillates around 5775 m/s for a frequency of 13.3 kHz, their wavelength is about 0.43 meter. According to the conclusions of the last mentioned work, these waves cannot reflect beyond the isobar 1 Mpa within the stress deformed areas. In the absence of *in situ* stress data measurements, we aimed at determining whether the representative parameters $K_1=Q_1/\sigma_v$ and $K_2=Q_2/\sigma_v$ with $Q_1=\sigma_1$ and $Q_2=\sigma_2$, $\sigma_v$ being the vertical stress, that is to say the lithostatic pressure, could provide values satisfying one of the following possibilities :
1) $(K_1+K_2)/2 = 0.7$, which represents a medium of weak or tensile stresses,
2) $(K_1+K_2)/2 = 1$, which represents a medium of constraint stresses,

3) $(K_1+K_2)/2 = 1.3$, which represents a medium of shear stresses.
The horizontal stress anisotropy is calculated with $(Q_1-Q_2)/[(Q_1+Q_2)/2]$.

Figure 5 shows four representative cases in function of depth. At 1000 meters (thickness of the stress deformed area : 0.40 meter) and 2100 meters (thickness of the stress deformed area : 0.55 meter), the case of $(K_1+K_2)/2=1.3$, that is to say of a shear domain, appears appropriate. At 1600 meters there are two cases corresponding to the presence of two thicknesses of the stress deformed areas : 0.55 and 0.40 meter. These are respectively the case of $(K_1+K_2)/2=1.3$ and the case of $(K_1+K_2)/2=1.075$ representative of a compressive domain. The anisotropy is very strong inside the shear domain (77 to 108 %) as well as inside the compressive domain (88 %), which is not surprising in an earthquake zone. The relative steadiness of the parameter $K_1$ within the shear domain with respect to depth, the value of which ranging from 1.80 to 2.00, backs up our interpretation. In order to research at 1600 meters of depth the kind of stress domain causing the deformed area thickness of 0.40 meter simultaneously to the shear domain, we have assumed that $K_2$ had to remain steady (=0.60) whatever the stress domain. As a result, the stress domain is compressive (the value of $(K_1+K_2)/2=1.075$ is close to 1.0) and the anisotropy is 88 %.

**Discussion : bayonet-shaped fault**

In order to integrate the contradictory effects of the simultaneity of shear and compressive domains at some depths, as well as the presence of shear zones at the depths 1150-1220 m, 1345-1420 m and 1820-1880 m (Boness and Zoback, 2004), the assumption of constraints acting on each side of a bayonet-shaped fault, as shown on Figure 6, allows us to have a coherent model of the San Andreas Fault in the vicinity of Parkfield. As a consequence, this fault would not be planar, at least in this location.

There are many arguments which strengthen this hypothesis.
- The epicentres of the quasi-permanent micro-seismicity appear almost entirely shared on the western side of the fault and on its surface trace, and almost never on its eastern side.
- Korneev et al. (2003) put into evidence along the San Andreas Fault the propagation of fault-zone guided waves within a 100-200 m wide zone at seismogenic depths and with 20-40 % lower shear wave velocity than the adjacent unfaulted rock. The authors think that this attenuation would be due to dewatering, but in fact the zigzags of the "bayonet" configuration are a mechanical argument which can better explain this phenomenon.
- Townend and Zoback (2004) show that within the San Andreas Fault zone the orientation of the maximum horizontal stress varies with depth and that the frictional coefficients, much larger than the typical 30-35° of strike-slip fault, reveal a fault-normal compression in some places. This ought to appear abnormal in the case of a planar vertical fault, but in fact our calculations gave values representative of compressive and shear zone. Besides, Chéry et al. (2004) point out the difficulties of a simple model for explaining data which appear contradictory in the case of a planar vertical fault, such as stress rotation with depth, effective friction and stress measurements made in both the far field and in the SAFOD pilot hole.

**Conclusion**

The calculation of the thicknesses of the stress modified areas around a vertical borehole, using body wave multiples, in order to estimate the horizontal stress domain, applied to the SAFOD Pilot hole, supplied surprising, but coherent, information. Probably because of the numerous breakouts which disturb the propagation of acoustic shear waves, only P waves supplied useful information. The depth 1270 meters separates an upper region of a steady thickness of the stress modified areas possibly corresponding to a shear domain, and downwards there are simultaneously two values of the thicknesses of the modified areas, particularly between 1500 and 1600 m of depth, able to correspond to a compressive and a shear domain.

The hypothesis that the San Andreas Fault could be bayonet-shaped instead of planar is able to explain not only the stress domain noticed above, but also the other recent available information in the literature about this fault, such as the presence of a fault zone of low shear wave velocity, stress rotating with depth, and the large angles of the frictional coefficients. Besides, this fault geometry could allow us to examine why the mechanical energy acting on the San Andreas Fault proves to be weak for a strike-slip fault.

**Acknowledgments**

I am grateful to M.D. Zoback and N.L. Boness (Stanford University) who provided to me the acoustic data.

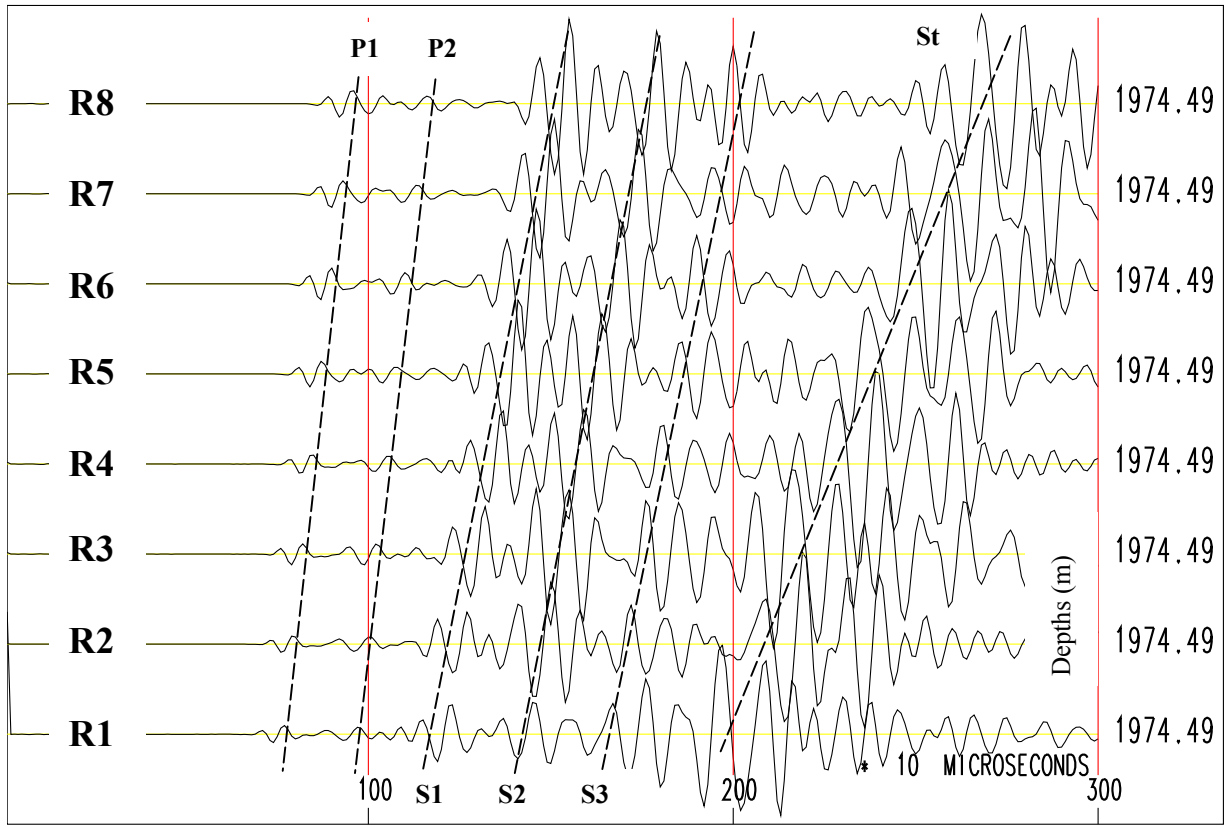

Figure 1 : 2D display of the move-out of the waveforms from the eight receivers R of the acoustic probe.
P = P wave    S = S wave    St= Stoneley wave
The depth on the right is indicated in meters

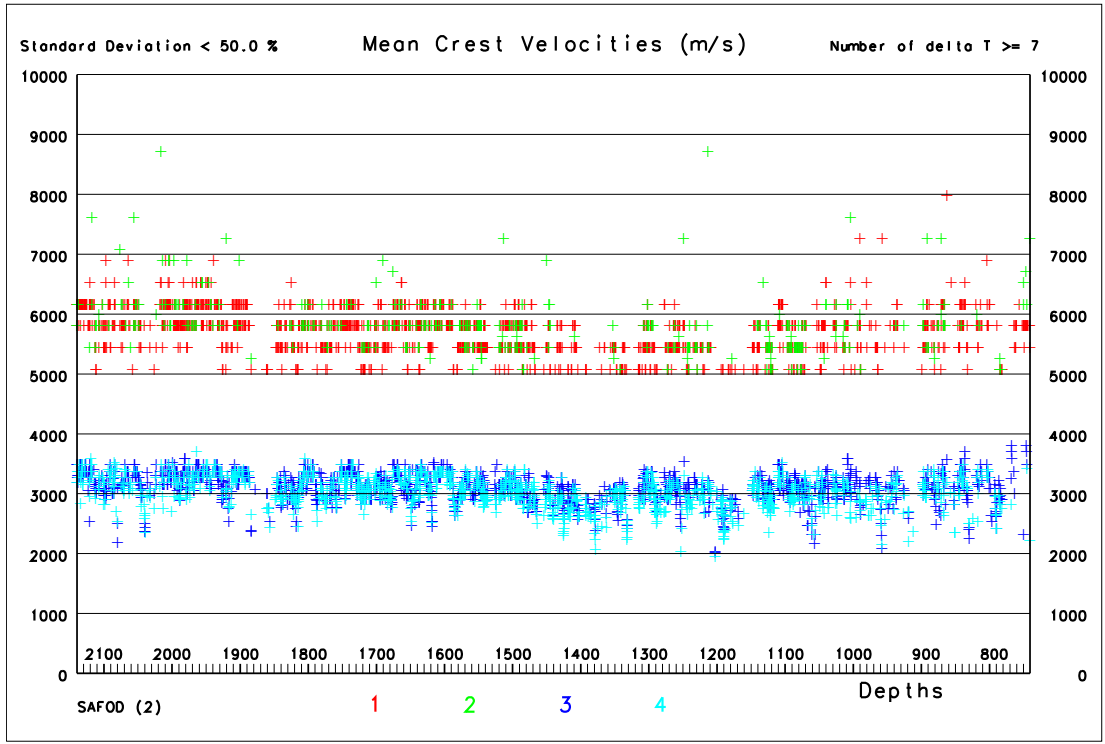

Fig. 2 : Velocities of :
**1 :** the first P wave arrival ; **2 :** the second P wave arrival
**3 :** the first S wave arrival ; **4 :** the second S wave arrival

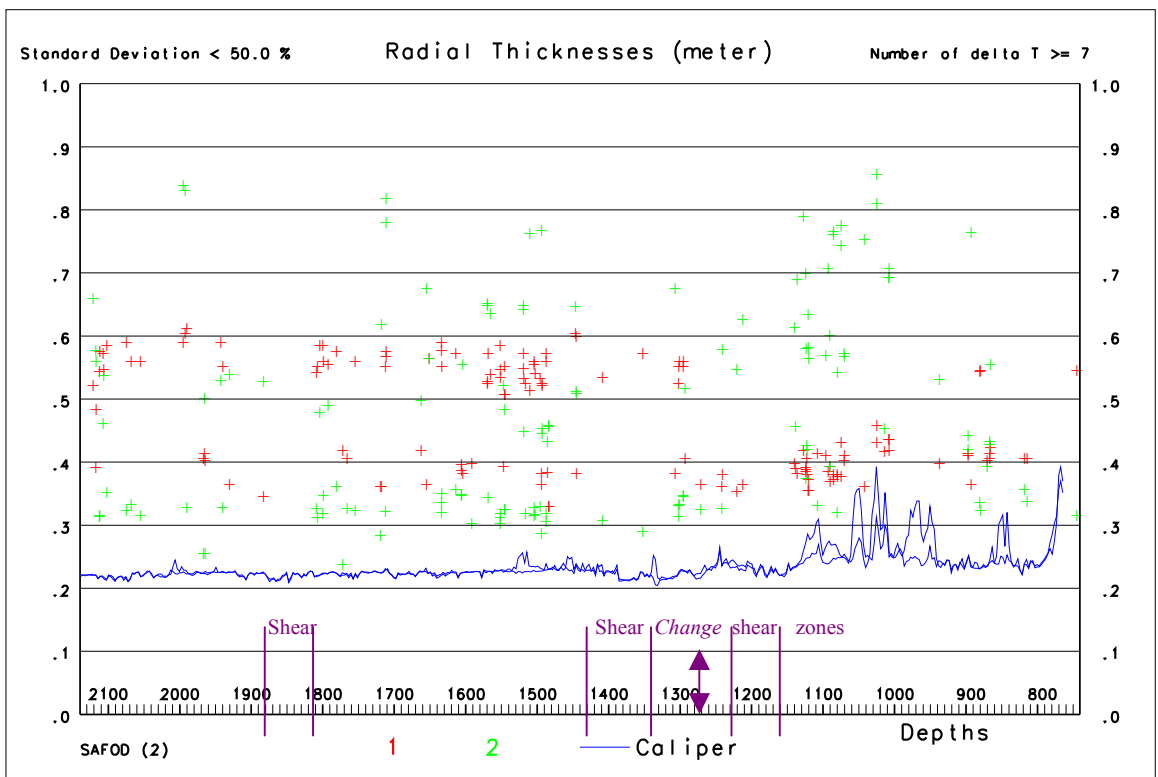

Fig. 3 : Radial thicknesses of the stress modified areas around the hole calculated in meters :
1 (red crosses) : from the P waves,
2 (green crosses) : from the S waves.

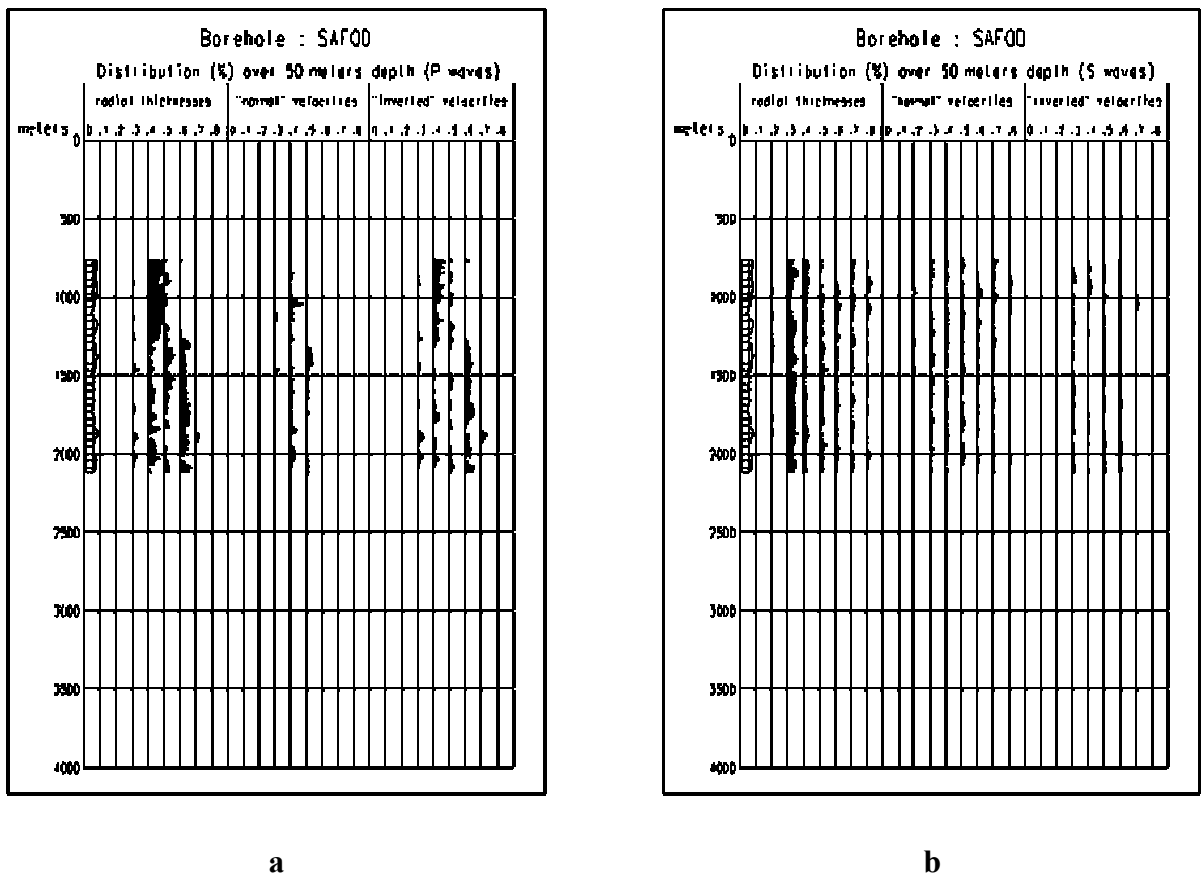

         **a**                               **b**

Figure 4 : Results calculated from
    a) P waves,
    b) S waves.

**1)** Distribution of the thicknesses of the stress modified areas around the well over 50 meter depth in respect of depth by grades of 0.1 meter. The width between two vertical lines represents 100 % of the corresponding size.
**2)** Percent of the "normal velocities" (the first arrival is faster than the second one).
**3)** Percent of the "inverted velocities" (the first arrival is slower than the second one).
*The squared area represents the percents of resonance over 50 meter depth in respect of depth.*

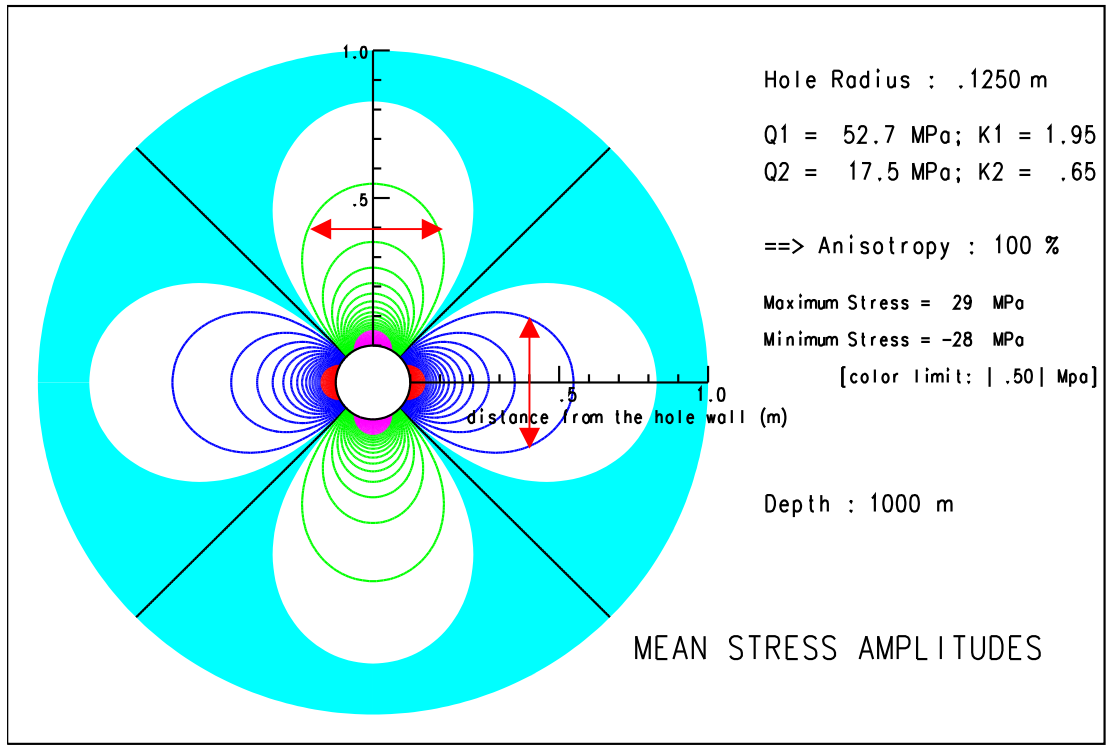

Figure 5a : Stress model able to correspond to the P wave results
at the depth of 1000 meters (*thickness : 0.40 m*)
**(K1+K2)/2 = 1.3**

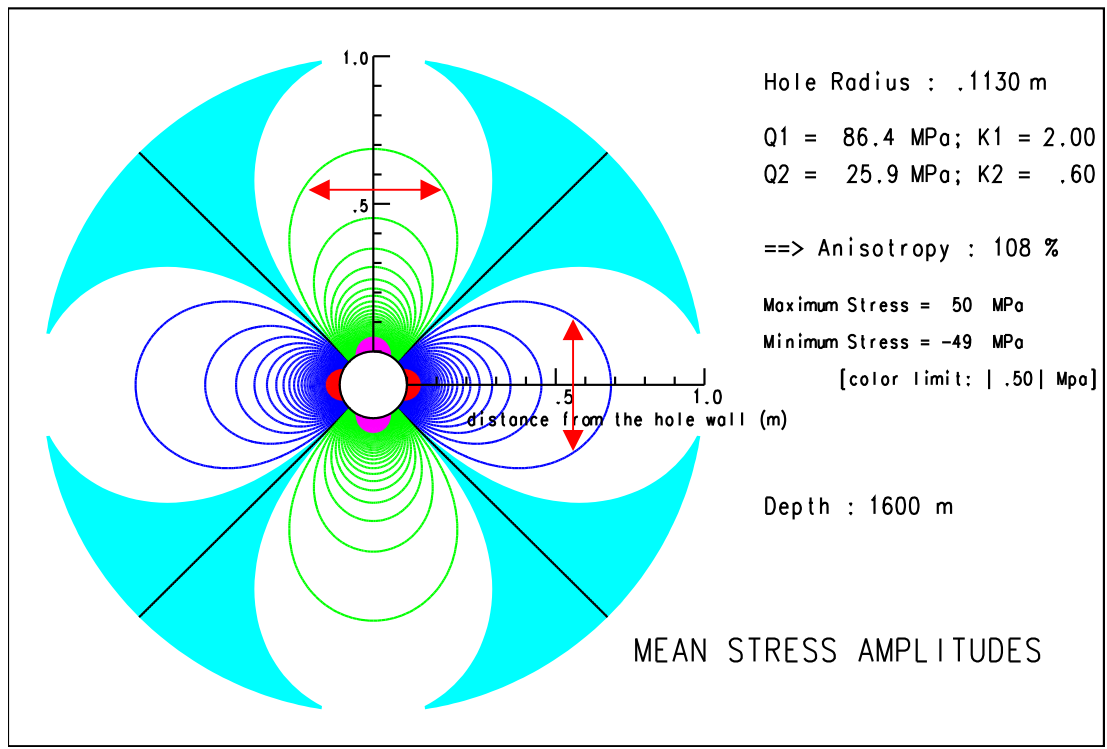

Figure 5b : First stress model able to correspond to the P wave results
at the depth of 1600 meters (*thickness :0 .55 m*)
**(K1+K2)/2 = 1.3**

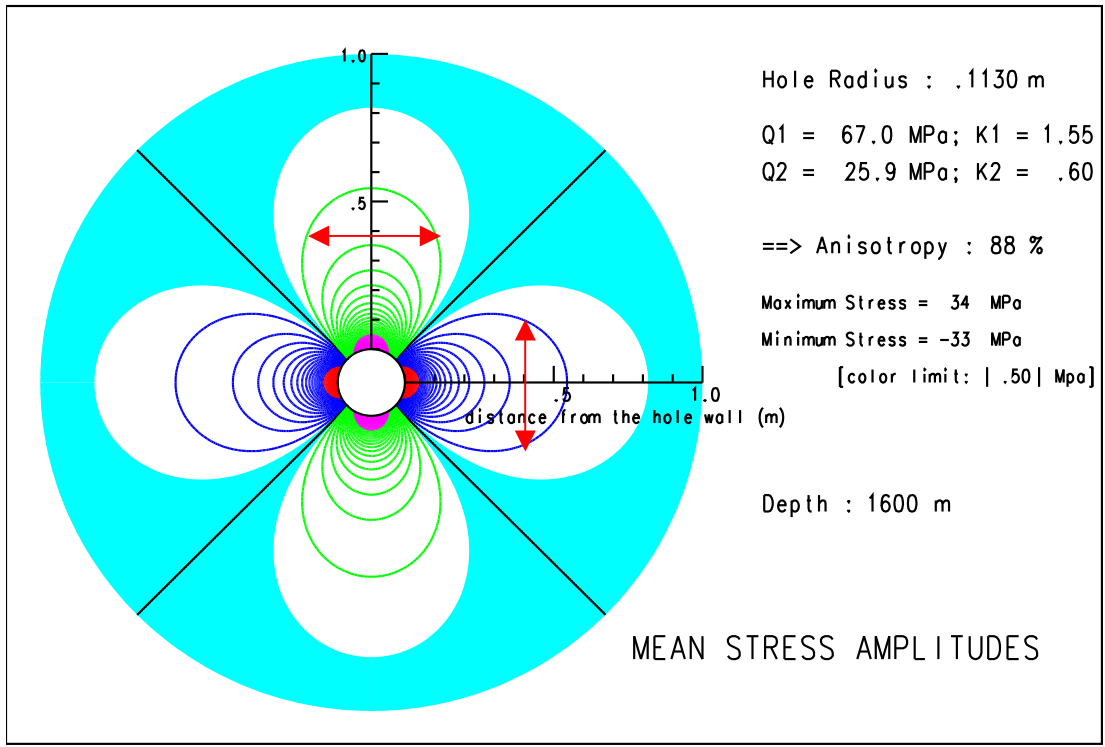

Figure 5c : Second stress model able to correspond to the P wave results
at the depth of 1600 meters (*thickness :0 .40 m*)
**(K1+K2)/2 = 1.075**

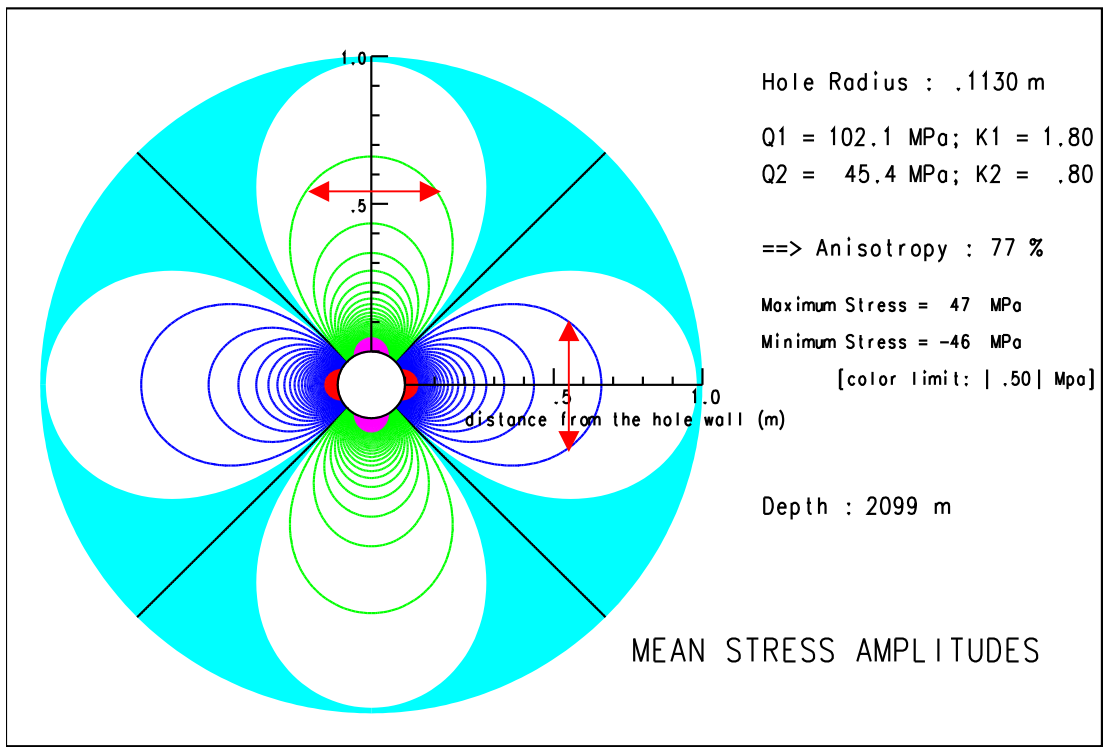

Figure 5d : Stress model able to correspond to the P wave results
at the depth of 2100 meters (*thickness :0 .55 m*)
**(K1+K2)/2 = 1.3**

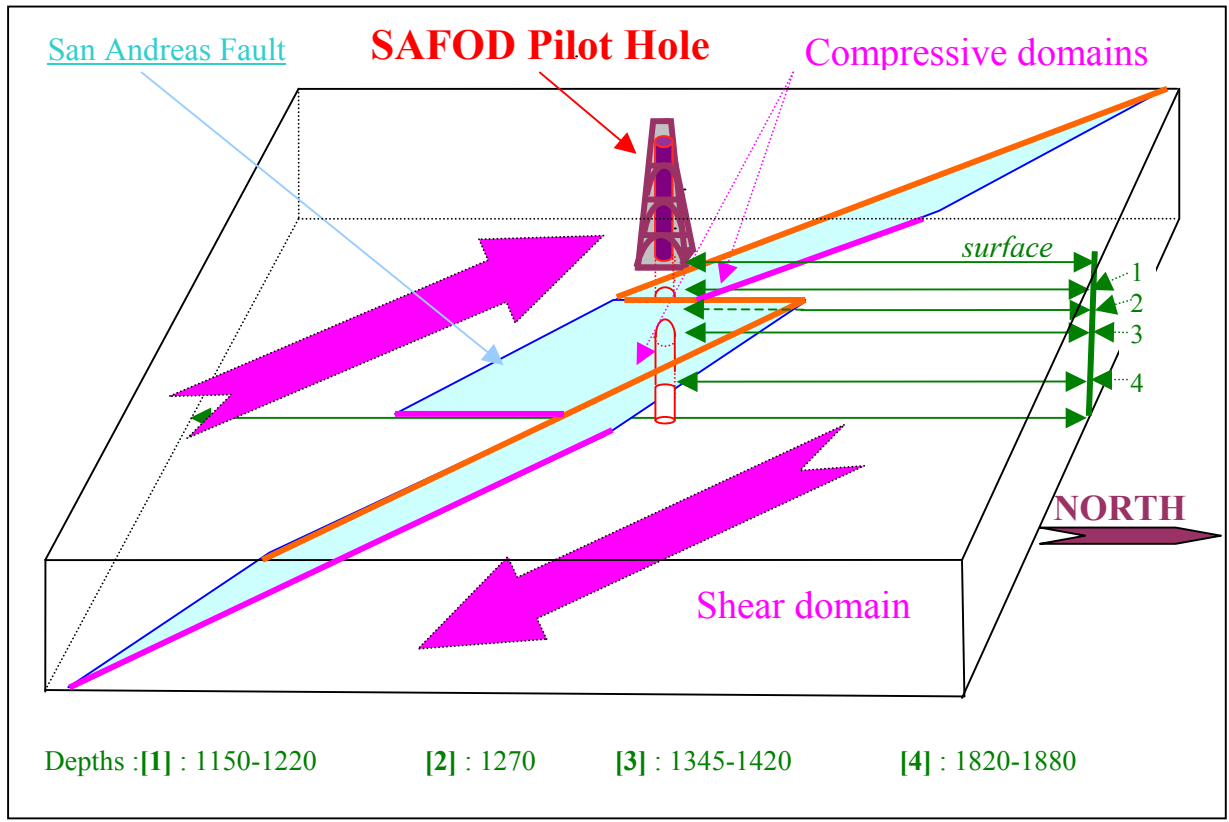

Figure 6 : Model of the fault crossing the SAFOD Pilot Hole.
The depths marked [1] [3] and [4] are the locations of shear zones.